\documentclass[conference]{IEEEtran}
\usepackage{cite}

\ifCLASSINFOpdf
\else
\fi
\usepackage{cleveref}

\usepackage{url}

\usepackage{graphicx}

\usepackage{subcaption}
\usepackage{tabu}

\usepackage{amsthm}
\usepackage{float}
\usepackage{algorithm}

\usepackage{algorithmic}
\usepackage[usenames,dvipsnames]{xcolor}

\usepackage{booktabs}

\begin{document}
%
\title{Gargoyle: A Network-based Insider Attack Resilient Framework for Organizations}

\author{\IEEEauthorblockN{Arash Shaghaghi\IEEEauthorrefmark{1}\IEEEauthorrefmark{2}, Salil S. Kanhere\IEEEauthorrefmark{1}, Mohamed Ali Kaafar\IEEEauthorrefmark{2}\IEEEauthorrefmark{4}, Elisa Bertino\IEEEauthorrefmark{4} and Sanjay Jha\IEEEauthorrefmark{1}}
\IEEEauthorblockA{\IEEEauthorrefmark{1}The University of New South Wales (UNSW Sydney), Australia\\
\IEEEauthorrefmark{4}Macquarie University, Australia \\
\IEEEauthorrefmark{3}Purdue University, USA \\
\IEEEauthorrefmark{2}CSIRO Data61 \\
\{a.shaghaghi, salil.kanhere, sanjay.jha\}@unsw.edu.au\\
dali.kaafar@mq.edu.au\\
bertino@purdue.edu}
}


%


\maketitle

\begin{abstract}
`Anytime, Anywhere' data access model has become a widespread IT policy in organizations making insider attacks even more complicated to model, predict and deter. Here, we propose Gargoyle, a network-based insider attack resilient framework against the most complex insider threats within a pervasive computing context. Compared to existing solutions, Gargoyle evaluates the trustworthiness of an access request context through a new set of contextual attributes called Network Context Attribute (NCA). NCAs are extracted from the network traffic and include information such as the user's device capabilities, security-level, current and prior interactions with other devices, network connection status, and suspicious online activities. Retrieving such information from the user's device and its integrated sensors are challenging in terms of device performance overheads, sensor costs, availability, reliability and trustworthiness. To address these issues, Gargoyle leverages the capabilities of Software-Defined Network (SDN) for both policy enforcement and implementation. In fact, Gargoyle's SDN App can interact with the network controller to create a `defense-in-depth' protection system. For instance, Gargoyle can automatically quarantine a suspicious data requestor in the enterprise network for further investigation or filter out an access request before engaging a data provider. Finally, instead of employing simplistic binary rules in access authorizations, Gargoyle incorporates Function-based Access Control (FBAC) and supports the customization of access policies into a set of functions (e.g., disabling copy, allowing print) depending on the perceived trustworthiness of the context.


\end{abstract}

%
\IEEEpeerreviewmaketitle

\section{Introduction}
The most secure organizations including U.S. intelligence agencies and the armed services have not been immune to insider threats. As discussed in \cite{bunn2017insider}, some of the well-known examples of insider attacks include 1) the former National Security Advisory Sandy Berger who removed highly classified documents from the National Archives to review them at his office; 2) John Deutch, the CIA director who handled highly sensitive classified information on an insecure computer connected to the Internet; and 3) the original WikiLeaks incident involving Bradley Manning who downloaded over 700,000 highly classified documents onto compact disks causing the largest leak of military and diplomatic cables in U.S. history.  \par 
The insider threat risk is much higher in smaller and less prepared organizations. In fact, Insider Threat Report 2018 \cite{insidertreport2018} reports on two key observations. First, ninety percent of participating organizations were vulnerable to insider attacks with the increased risk being associated with the growing number of mobile devices with sensitive access to data and users with excessive privileges. Second, organizations have reported shifting their focus on detection of insider threats and deterrence methods but the majority of these are still at the early stages of developing their insider threat program. \par
The main limitations of existing literature can be summarized as follows. First, access control systems fail to swiftly adapt to the `negative changes' in user's behavior even if it suggests attacking the system \cite{baracaldo2017g}. The cause for this is that the technical precursors used to predict insider attacks such as download and use of hacker tools, unauthorized access to systems and resources, system access after task termination, and inappropriate Internet access are not included in the access control process \cite{moore2008big}. In fact, with existing solutions, the system's trust is established independently of these actions.
Second, assuming that users follow the security rules. For example, trusting an employee to access confidential information in a secure room over a secure connection and in the presence of a supervisor rather than having an access control system enforcing this policy. In other words, adaptive access control solutions should model context using a range of attributes required to deter insiders. The third limitation is trusting the user's device integrated sensors for retrieving the context attributes. These sensors are within the user premises and hard to protect against attacks. In fact, alternative sensors are required to, at least, validate such information. Fourth, a binary approach to access decisions is not suitable for pervasive computing context. For instance, a user requesting access to objects in a less trusted environment (e.g., when in the proximity of flagged users, using an insecure WiFi connection), should not be allowed to share information but still be able to fulfill the assigned task. For instance, this could be achieved by disabling functions such as Print and Email for that access and allowing View and Search. The fifth limitation is relying on single protection measures. In fact, there need to be alternative backup access control enforcement points that function independently of the user's device. Hence, if an attacker compromises a mobile device, it can still be prevented from targeting the organization's services. \par
To address the aforementioned limitations, we designed and developed Gargoyle. 
Compared to existing solutions, Gargoyle is designed against the most complex malicious insider threats and aims to detect and deter an insider throughout its key attack phases (see \S\ref{background}). Gargoyle includes a new set of attributes for context analysis called Network Context Attribute (NCA). NCAs are extracted from the device generated network traffic and include information such as the user's device capabilities, security-level, network connection type, network status, current and prior interactions with other devices, and suspicious online activities. For instance, Gargoyle detects devices equipped with hacking tools (e.g., port scan, vulnerability scanners) or connected over a suspicious network point (e.g., Intrusion Detection System (IDS) raising the alarm for certain network segments). It can also detect devices with outdated software, unusual behavior (e.g., unusual locations, interactions with devices, etc.) and suspicious browsing history such as accessing blacklisted domains. Gargoyle leverages the capabilities of Software-Defined Network (SDN) and retrieves contextual information by passively analyzing network traffic. This enables Gargoyle to function independent of the user's device integrated sensors and be portable to different organizations deploying SDN with ease. \par
Gargoyle assesses the risk associated with an access request through NCAs and by modeling the user's behavior (both current and historical). Compared to the existing solutions, Gargoyle can be programmed to apply access restrictions both at host-level and network-level. In fact, Gargoyle's SDN App (GSDN) enhances policy enforcement and facilitates a defense-in-depth protection model. For instance, a suspicious device can be restricted from accessing organization's network until further investigations. Finally, by implementing Function-based Access Control (FBAC)\cite{desmedt2016function}, Gargoyle's mobile App (GAPP) can restrict a set of functions for a data requestor depending perceived trustworthiness of a context. \par

The rest of this manuscript is structured as follows. We review key background information in \S\ref{background}. Thereafter, the threat model that Gargoyle is designed against is presented in \S\ref{threatmodel}. An overview of our proposed solution is presented in \S\ref{overview}, followed by discussion of its main components, context extraction (\S\ref{contextextraction}) and access control (\S\ref{accesscontrol}). We present our implementation (\S\ref{implementation}) and evaluation of Gargoyle (\S\ref{evaluation}) followed by a discussion of related work in \S\ref{relatedwork}. We conclude the paper outlining the future work in \S\ref{conclusion}.


\section{Background} \label{background}

\subsection{Software-Defined Network (SDN)}
Software Defined Networking (SDN) separates the control plane of the network from the data plane. It facilitates network programmability and grants the ability to manage, amend and control the network behavior dynamically. SDN enables centralized control of data plane forwarding devices independent of the technology used to connect the devices while maintaining live and centralized network-wide view of all the data path elements. One of the key practical benefits of SDN comes with its Application Plane, allowing the development of new services that leverage the added network-layer capabilities. In fact, many different SDN applications have already been proposed, and the the current focus is to have an App Store support, where customers can dynamically download and install network apps \cite{arashsurvey}. \par

Major service providers and ISPs have already adopted SDN at their data centers. Indeed, deploying SDN has many benefits for organizations including networking infrastructure cost reduction, simpler management, reduced complexity, improved security, more flexibility and better support for innovation \cite{kreutz2015software}. Hence, it is expected that organizations of all sizes have no choice but to gradually shift towards SDN adoption \cite{riverbed}. 


\subsection{Function-based Access Control (FBAC)}
Function-based Access Control (FBAC) \cite{desmedt2016function} avoids a binary approach in access authorizations. With FBAC, access authorizations are no longer stored as a two-dimensional Access Control Matrix (ACM). Instead, FBAC stores access authorizations as a three-dimensional tensor (called Access Control Tensor). Hence, applications no longer have blindfolded execution right, and users can only invoke commands that have been authorized for each object. In practice, objects are data blocks and functions are the commands available in applications, such as Copy/Paste, Email, and Print. For deployment, FBAC can be implemented as either restricting a set of functions or allowing them for the data objects. 

\subsection{Insider Attack Phases}
Based on \cite{liu2018detecting}, we assume a malicious insider goes through the following four main phases to deliver attacks against organization data and its systems. The `Reconnaissance' phase involves an attacker exploring the victim environment's computer systems, networks and applications for vulnerabilities through means such as port scan, network, web application and database vulnerability scans. The `Delivery' phase involves engaging the victim environment through actions such as social engineering to deliver a seemingly innocuous communication but coupled with a malicious payload such as remote access trojans, rootkit backdoors, and keyloggers. After that, during the `Exploit and Install' phase, the insider aims to escalate his privileges and attempts to install trojans and backdoors. These are then used in the `Command \& Control' phase to launch various attacks such as botnet, Denial of Service (DoS) and email spam. 



\section{Threat Model} \label{threatmodel}
\begin{figure*}[!htb]
  \centering	
  \includegraphics[width=6.5in, scale=0.50]{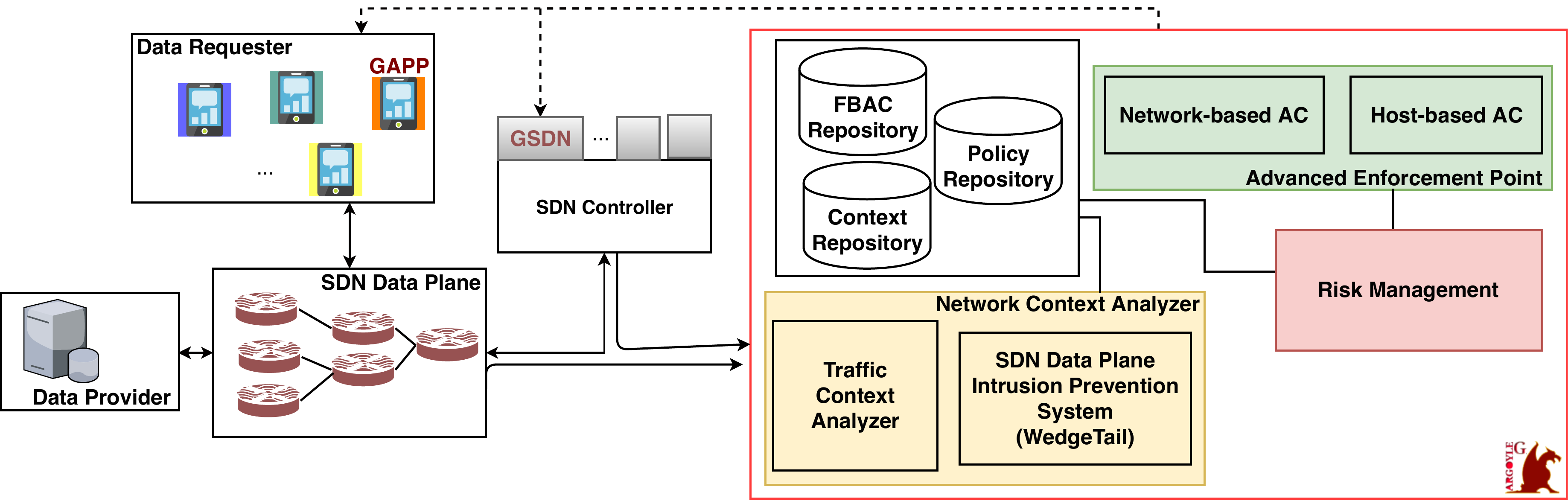}
  \caption{Gargoyle's Architecture: the three main components are shown in coloured boxes.}
  \label{architecture}
\end{figure*}

Compared to existing solutions (see \S\ref{relatedwork} for related work), Gargoyle is designed against the most advanced cases of insider attack within organizations adopting a Bring Your Own Device (BYOD) policy. Here, we assume an insider placed within the organization's boundaries who is equipped with portable computing devices (e.g., mobile phone, laptop, etc.). We assume the insider and all the other data requesters access files only using Gargoyle's mobile application (i.e. files are only available when requested through the App). All data requestors and providers are assumed to be exchanging data through either wired or wireless network connections. In other words, other data exchange services such as Bluetooth function cannot be used to exchange files between the data provider and requester. \par
We assume that the insider's goal fits into one of the three common insider attack categories, including: I) data exfiltration, II) data integrity and availability breach and III) ICT systems sabotage \cite{team2014unintentional}. He may achieve these goals either by violating his own access privileges or attacking other users who are retrieving sensitive information. In fact, the insider may be equipped with any other device equipped with software used to hack other devices and organization services. The insider would typically have access to networking functions as afforded to a typical employee which could include unrestricted Internet access, ability to connect to organizational services and and browsing the Internet. \par
The insider may be capable of performing attacks such as social engineering to trick a legitimate user to request access to data within an insecure context. For instance, to run a data exfiltration attack, the attacker might exploit known router vulnerabilities and gain unrestricted access to a set of forwarding devices and tamper with traffic being routed. In fact, although Gargoyle is a network-enabled service, we do not assume the data plane devices to be necessarily secure given the range of attacks and vulnerabilities reported against the routing devices \cite{shaghaghi2017wedgetail}. Instead, we only assume the network control plane (or, the `network brain') to be secure -- this includes Gargoyle's SDN Application (GSDN) installed a top of the controller. In fact, this is a reasonable assumption given that compromising the core network infrastructure components, which is typically placed away from the user boundary is a much harder target for an adversary. Moreover, with Software-Defined Network (SDN), which Gargoyle is built on, a compromised controller means a compromised network altogether, which is a threat to all organization services at all levels. In fact, most, if not all, SDN-based services are built assuming that the underlying technology is secure at both control and data plane \cite{ali2015survey, scott2016survey}.
Finally, it is assumed all organization files are stored in the Atomic format required for FBAC and information is accessible to data providers only through GAPP. Moreover, it is assumed that GAPP is trustworthy and capable of connecting with other Gargoyle modules securely (e.g. using TLS).


\section{Overview of Gargoyle} \label{overview}
Gargoyle's architecture is inspired by Crampton and Huth's `extended access-control architecture' \cite{crampton2010towards} and integrates the observations of context and the assessment of risk into the access control mechanism. As shown in Figure 1, the proposed solution has three main components: (i) Network Context Analyzer, (ii) Risk Management, and (iii) Advanced Enforcement Point (AEP). The Network Context Analyzer component extracts context information pertinent to each user both over-time and in real-time. This information (called Network Context Attribute, or NCA) is retrieved by analyzing the network traffic collected for user devices using the `Traffic Context Analyzer' module. Gargoyle is a network-based solution relying on forwarding devices for context extraction and access enforcement, and therefore, it has to ensure that the data plane forwarding devices have not been compromised. To achieve this, it integrates the reports from the SDN Data Plane Intrusion Prevention System (IPS) when evaluating the context trustworthiness.  In fact, for Gargoyle to decide the most appropriate access limitation (e.g., whether to disable certain functions or block all access requests originated from certain network zones), the IPS should be capable of identifying the compromised forwarding devices, locating them and detecting their specific malicious actions (e.g packet fabrication, forwarding, etc.). Here, we integrate our earlier work WedgeTail \cite{shaghaghi2017wedgetail} into Gargoyle's design to satisfy this requirement. WedgeTail is a controller-agnostic Intrusion Prevention System (IPS) designed to ‘hunt’ for forwarding devices failing to process packets as expected. It tracks packet paths when traversing the network and generates their corresponding trajectories. Thereafter, by comparing the actual packet trajectories with the expected ones, WedgeTail detects malicious forwarding devices, locate them and identify the specific malicious actions. \par

Finally, the Risk Management component according to the policies specified by the Policy Repository (PR) and FBAC Repository forwards a set of access authorizations to the AEP component. AEP's instructions include actions for the `Host-based' and `Network-based' access control modules. The host-level access control involves allowing or restricting a set of functions for data objects, which are enforced by Gargoyle's mobile Application (GAPP). The network-based access control module implements a set of restrictions at network-level through Gargoyle's SDN Application (GSDN) -- these are completely independent of the host-level restrictions and have a much higher granularity level. In other words, network-level access enforcements are not applied at file level or functions, but instead apply access restrictions such as disconnecting the device from the network altogether. Network-level access enforcement is mostly relevant during the later phases of insider attack such as `Command \& Control" and when the attacker goal is to sabotage the enterprise ICT systems.

\section{Gargoyle's Context Extraction} \label{contextextraction}
\begin{figure}
  \centering
\includegraphics[height=2.2in, width=2.8in]{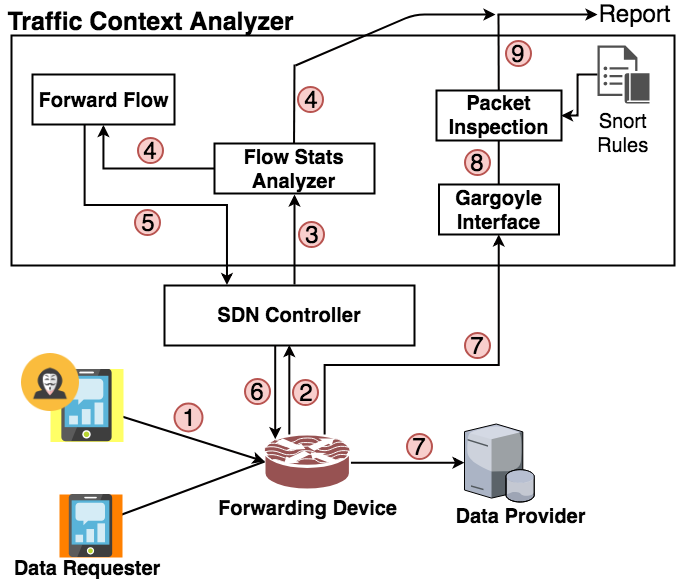}
 \caption{Gargoyle's Traffic Context Analyzer}
\label{Figure1:Setup}
\end{figure}
As mentioned, Gargoyle's Network Context Analyzer component is responsible for extracting the Network Context Attributes as well as the data plane security status. As mentioned, we rely on WedgeTail \cite{shaghaghi2017wedgetail} to inspect the forwarding devices and report on malicious devices and their specific actions to Context Repository. Here, we, therefore, focus on the Traffic Context Analyzer component uses to retrieve the NCAs. \par
There are no restrictions on what an NCA can represent. In fact, NCA may be any information that can be systematically and reliably extracted from the network traffic. In this work, we define the following as attributes of interest: a) the user's device capabilities including specific operating systems and tools; 2) security-level that depends on current and prior interactions, location and current user ID; 3) current and prior interactions with other devices; 4) network connection status including medium of communication (i.e. wired or wireless); and 5) suspicious online activities including traffic directed to restricted domains, IPs and services -- refer to the sample insider scenarios explained in \S\ref{evaluation} for examples. \par
In order to extract NCAs we take a systematic approach that can be imported to different networks. Figure 2 illustrates the architecture of our solution, which we implemented over Floodlight controller. The user traffic through either wired or wireless connection is routed through a forwarding device (1). As per OpenFlow standard, the first packet of the flow is sent through to the controller (2). At this point, the Flow Stats Analyzer collects the network statistics to detect possible anomalies including DoS attacks (3). This information is continuously updated and forwarded to the Risk Management component for each user's traffic being routed over each port. At this point, the Forward Flow decides whether the flow needs to through packet inspection as well as being forwarded to the Data Provider. If so, the controller install the corresponding flow rule at the forwarding device (5). Note that packet headers are not forwarded to the controller and this is required so that the Traffic Context Analyzer receives a copy of network traffic (7) for inspection as well. At this point, packets are inspected and any NCA is extracted (8) and attached to the report being forwarded to the Context Repository (9). \par
A common attribute to evaluate user context is location, which we define as another type of NCA. In this case, we adopt the solution developed in our earlier work \cite{shaghaghi2016towards}. In fact, with OpenFlow, whenever a packet is received by a forwarding device ($FD$), and it does not match any of its existing forwarding rules then a $packet_{in}$ message containing the $FD_{id}$ and $Port_{id}$ is sent to the governing controller. The controller uses this information and creates a dynamic geo-location lookup table. This table matches the user’s device IP with a forwarding device port. The network locations retrieved through $FD_{id}$ can be matched to different sections within the building. For example, in Figure 4, $Zone 1$ is associated to access point $R1$. An issue to consider for wireless devices is managing the signal coverage. This can be solved using proper and careful positioning of access points and signal blocking solutions \cite{coleman2010certified}.

\section{Gargoyel's Access Control} \label{accesscontrol}
\begin{figure}
  \centering
\includegraphics[height=2.2in, width=2.8in]{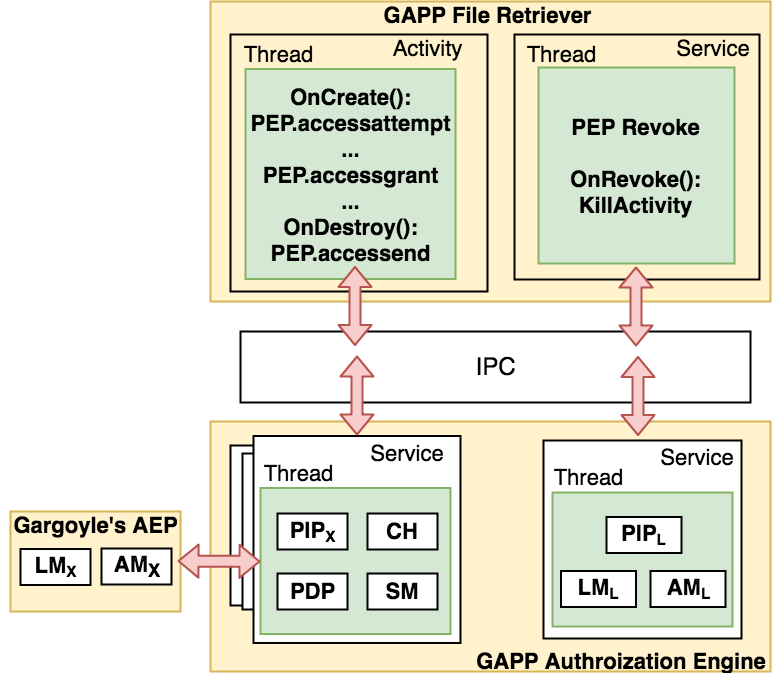}
 \caption{GAPP implementation in Android}
\label{Figure1:Setup}
\end{figure}
Gargoyle's Access Control component is composed of two main modules: Risk Management and Advanced Enforcement Point (AEP). When an access request is retrieved, the risk management module queries the context repository for all recent and historical NCAs and network status reports for the user. It then retrieves the user policies from the Policy Repository along. For instance, the organization default RBAC model may define user A as an employee with access to sensitive records (See Sample Scenario-I in \S\ref{evaluation}). At this point, it queries for contextual FBAC policies from the FBAC repository. This repository specifies the functions that should be allowed for each context. For instance, it specifies that if a user is connected to the network over wireless and is requesting access to sensitive documents, then Copy function must be disabled. Hence, any function not restricted is assumed to be enabled. Here, we use XACML \cite{rissanen2013extensible} to store the contextual FBAC policies. \par
The risk management submits a set of access restrictions to AEP. We call it Advanced Enforcement Point since unlike existing solutions Gargoyle policy enforcement is applied both at the host-level and network-level. For host-level, the granularity is much higher but critical in cases where the insider may be aiming to sabotage ICT systems. For instance, if the flow rate indicates a DoS attack, the user is quarantined from the network. We adopt our earlier system developed in PEPS\cite{shaghaghi2016towards} for implementing the host-level access restrictions. \par
With Gargoyle, the host-level access restrictions are set up dynamically and the level of functions for each data segments. For this purposes, we developed Gargoyle's Android application (GAPP). GAPP's design is inspired by \cite{lazouski2017stateful} and Figure 3 shows its architecture along with the main functions. Context Handler (CH) is the front end and manages the communication with PEP. Policy Decision Point (PDP) evaluates the security policies and produces the decisions. Session Manager (SM) keeps track of the ongoing usage session to allow the continued enforcement of the policy. The Attribute Managers (AM) manage the retrieval and update of NCAs -- $AM_{L}$ refers to local attributes and $AM_{X}$ to remote ones. Policy Implementation Points (PIP) provide interfaces to query the AMs for retrieving and updating attributes -- $PIP_{L}$ refers to local attributes and $PIP_{X}$ to remote ones. Lock Manager (LM) guarantees consistency in concurrent retrieval and updating of NCAs.
 



\section{Implementation} \label{implementation}
We envision Gargoyle to be implemented as an application for SDN controllers. However, at this stage, to demonstrate Gargoyle's compatibility with different platforms and evaluate it over different controllers we implemented the Network Context Analyzer and Risk Management components as a proxy service sitting in between the control and data plane -- a similar approach is taken in solutions requiring advanced traffic analysis such as \cite{shaghaghi2017wedgetail, dhawan2015sphinx}. \par 
We developed GSDN as an application for Floodlight controller. Furthermore, we replicated our earlier work in PEPS\cite{shaghaghi2016towards} and setup MariaDB \cite{mariadb} as a data provider, which through our purpose built extension communicates with GSDN and defines dynamic network-level access control rules. To implement the host-based access control, however, we extended our earlier implementation of FBAC \cite{desmedt2016function} and developed GAPP. Currently, GAPP is an Android application capable of applying function restrictions (e.g., Copy/Paste, Email, Print) for data objects. Specifically, whenever an access request is received, the Risk Management component depending on the perceived trustworthiness of a context, instructs GAPP to disable a set of functions for different file segments (i.e., all functions enabled by default). 

\section{Evaluation} \label{evaluation}
\textbf{Overview} We evaluated Gargoyle over one thousand different simulated insider attack scenarios, which vary in terms of the number of active users, policies and attacks as well as the underlying organization setup including the physical map, networking interfaces, and configurations. We evaluated Gargoyle's performance in extracting context information and making access decisions. First, we collected actual network traffic within an organization setting, simulated an SDN network and re-routed traffic over it. At this point, according to the scenarios specified for evaluating Gargoyle's access control components (i.e., Risk Management and AEP), matching synthetic and curated traffic was injected through the network -- refer to Insider Scenarios for more details). In order to have a baseline to compare the performance of Gargoyle, we ran the same insider scenarios when having Role-based Access Control (RBAC) \cite{ferraiolo2001proposed}, solution proposed in \cite{lazouski2017stateful} and FBAC \cite{desmedt2016function}. We chose RBAC given that it is the most common access control model used in organizations. In particular we chose \cite{ferraiolo2001proposed} since it is based on UCON model \cite{park2004ucon} and is similar to our work in terms of motivation and threat model. Third, we evaluated the attack scenarios when having the default FBAC in-place -- which applies function restrictions based on user-roles. Finally, we also report on Gargoyle's mobile App (GAPP) prototype in terms of its performance and efficacy in applying access restrictions specified by AEP. \par 

\textbf{Network Setup} We setup a Mininet network comprised of up to 4 core forwarding devices (representing WiFi and Wireless network points), eight edge forwarding devices and a random number of hosts -- corresponding to the trace file being used. We hosted the simulated network on a machine equipped with Intel Core i5, 2.66 GHz quad-core CPU and 16 GB of RAM. We attached this network to a Floodlight controller equipped with GSDN and WedgeTail, which was running separately over an Intel Core i7, 2.66 GHz quad-core CPU and 8 GB of RAM. The same machine also ran other components of Gargoyle. \par

\textbf{Traffic Collection} We collected network traffic over 60 days period between 9 AM - 5 PM at a specific floor and building at UNSW Sydney. For this, all users' network traffic (wired or wireless) sent through the floor's switches were replayed to a specific interface accessible to us. The collected traffic included 94 identifiable users of whom 28 were present in more than 80\% of the data collection period. The collected traffic is split into batches of two hours. The traffic collected was sanitized using custom-written scripts, and all traffic not useful for NCAs were removed (e.g., traffic generated for services such as Dropbox LAN synch and Microsoft Windows update). \par  

\textbf{Replaying Traffic} We use Tcpreplay toolset \cite{tcpreplay} and custom written scripts for this. Tcpreplay uses tcprewrite to remap the source and destination MAC and IP addresses in the collected trace. We then regenerate the traffic with proper timing and replay over in the simulated network. For forwarding device flow entries, we created an interface for a subset of prefix found in a full BGP table from Route Views \cite{routeviews} and spread them randomly and uniformly to each router as ‘local prefixes’. We then computed forwarding tables using shortest path routing. 

\begin{figure}
  \centering
\includegraphics[height=2.5in, width=3.4in]{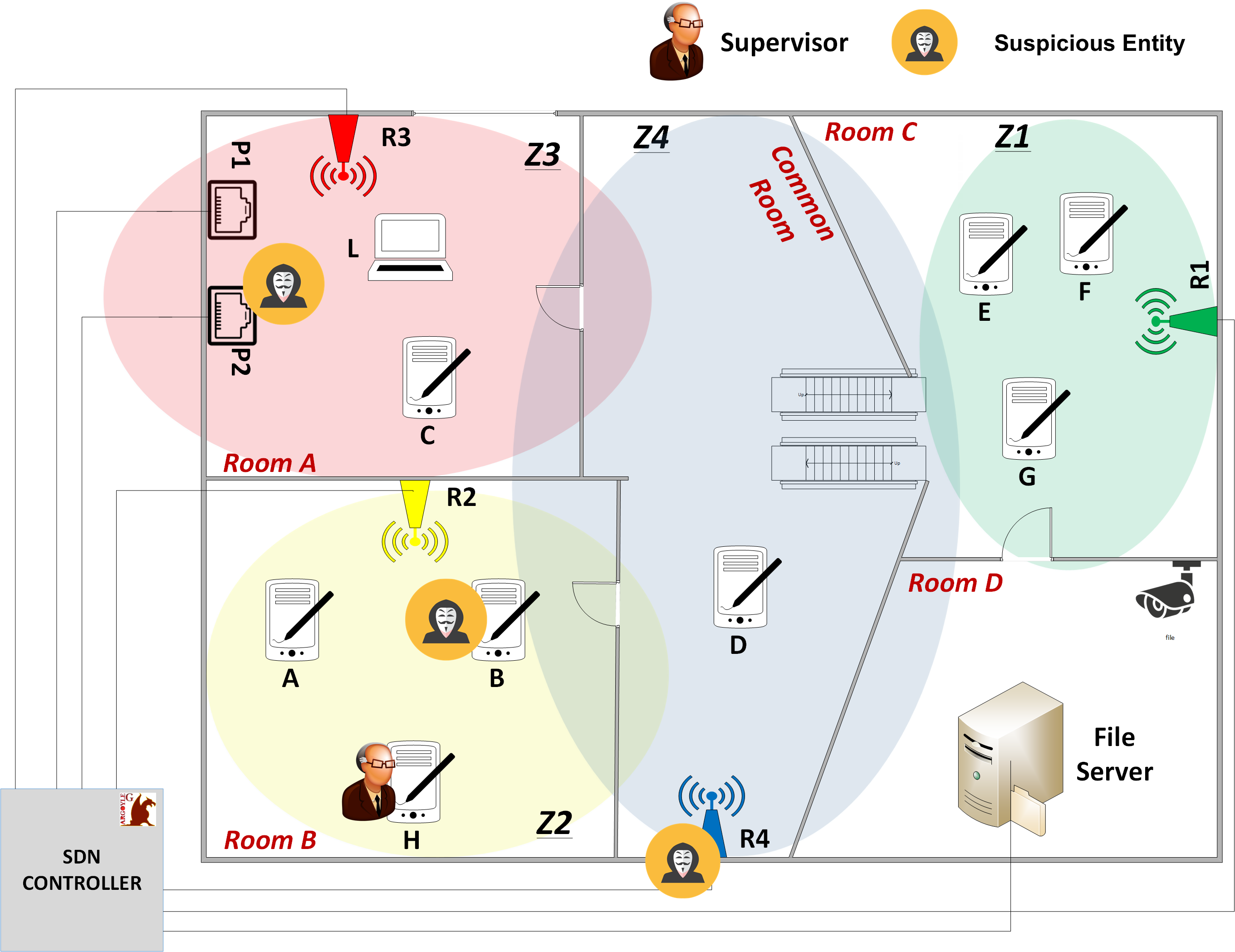}
 \caption{A sample insider scenario used to evaluate Gargoyle. For clarity, figure shows part of the actual map used during evaluation. Each coloured oval indicates the matching colour access point coverage.}
\label{Figure1:Setup}
\end{figure}

\textbf{Insider Scenarios} We defined seven different organizations each with a different map layout and network setup. Figure \ref{Figure1:Setup} shows part of one of the organization maps used for evaluating Gargoyle (simplified for clarity). The layouts had different network setup with a maximum of 4 core forwarding device and eight edge forwarding devices, which the network hosts associated with. In order to evaluate Gargoyle we implanted insider threat scenarios representing the four main phases of an insider attack (see \S\ref{background} and Sample Scenarios). Moreover, to further resemble real-word threats and evaluate Gargoyle's capabilities, we defined four categories of insider scenarios. In the first category, the data requester's (DR) device was malicious. To implement this, the custom script injected traffic indicating activities such as port scan in the DR's traffic or access to certain restricted websites. Similarly, for the second category, the traffic pertaining to a user in the proximity of a data requester was injected with signatures indicating suspicious activity (e.g., malware signature, a device equipped with Kali Linux, etc.). For the third category, we implanted malicious forwarding devices in the network both at core and edge, which data requester's traffic (either inbound or outbound to data provider) were routed through. The fourth category included compound cases, which involved one or more of the aforementioned scenarios (i.e., scenarios 1,2 and 3). Table II shows the distribution of different insider scenarios types. On average per each simulated scenarios, 35 users were present (with the maximum of 90), and we evaluated about 140 different scenarios for each organization map (see Table I for a summary). 


\begin{table}[H]
\center
\begin{tabular} {c|c}
    \toprule
       
    Number of different organization maps & 7  \\
    Number of different network configurations & 7  \\
    Number of edge forwarding devices (fixed) & 8 \\
    Maximum number of core forwarding devices & 4 \\
    Average number of users present in each scenario & 35  \\
    Average number of insider scenarios evaluated per map & 140 \\ \bottomrule
  \end{tabular}
\label{scenarios}
\caption{Experimental settings}
\end{table}

\begin{table}[H]
\center
\begin{tabular} {c|c|c}
    \toprule
       
   1 & DR's device was suspicious & 20\%  \\
   2 & Device in DR's proximity was suspicious & 30\%  \\
   3 & Network forwarding device was compromised & 10\% \\
   4 & Compound (e.g. both 1 and 2) & 40\% \\ \bottomrule
 
  \end{tabular}
    \label{attacktypes}
\caption{Approximate distribution of insider scenario types}
\end{table}

\begin{table}[H]
\center
\begin{tabular} {c|c}
    \toprule
       
	Requests blocked due to current suspicious behaviour & 184\\ 
    Requests blocked due to historic suspicious behaviour & 243 \\
    Requests blocked due to compromised forwarding devices & 58 \\
    Access granted with limited number of functions & 515 \\ \bottomrule
 
  \end{tabular}
    \label{attacktypes}
\caption{Stats on access authorizations}
\end{table}

\textbf{Sample Scenario-I} Consider Figure \ref{Figure1:Setup}. Device $D_A$ belonging to user $U_A$ with role $R_2$ connected to the network through access point $R2$ sends an access request to the File Server for File $F_1$, which contains sensitive paragraphs on the use of drones in war.  The organization policy $P_{F1}$ requires $F_1$ to be accessible to $R_i$ with $1<i<10$. It also states information to be accessible when the user's traffic is routed through the lower-level floors (i.e., Room A, B and Common Room). Moreover, it requires that all paragraphs flagged as `top secret' to be available for Print and Email only when $D_A$ is within a completely safe context (i.e., no threats are detected) and a supervisor is within the same wireless coverage zone. A generic organization policy $GP_1$ requires that if any device with hacking capabilities are detected in the proximity of a user requesting access to war-related documents, both users are immediately blacklisted and only allowed to access external services through Room C.  \par
During the simulation, first Device $B$ does not exist in $Z2$, and all requirements are met for $A$ to access $F1$. Hence, access is granted at time $t_1$ with GAPP granting Email and Print functions for the user. However, at $t_2$, device $B$ owned by user $U_B$ is added to $Z2$. There is no policy preventing access if $U_B$ is in the proximity of a data requester for $F1$. However, at time $t_2$, we inject traffic for $B$ that indicates it is equipped with Kali Linux and running UDP/NULL port scan. Upon detection of this, Gargoyle reacts according to the policies specified. 

\textbf{Sample Scenario-II} In the same figure consider device $L$ connected to the network through ethernet port $P2$ and wireless interface $R3$. This device is requesting access to $F1$ as well. Specifically, at time $t_3$, the device is requesting access through wired and at time $t_4$ through a wireless connection. A generic organization policy $GP_2$ specifies that if user's traffic is routed by any forwarding device flagged as suspicious or comprised, then all access should be blocked unless traffic can be routed through a safe path.  
Similar to sample scenario-I, $L$ is granted access to $F1$ at time $t_1$, where the context is retrieved as safe. However, at time $t_2$, $P2$ delays packets being routed. The malicious forwarding device along with the action is detected by WedgeTail and reported to Gargoyle. At this point, if $L$ is connected to the network through $P2$, then the access is dropped, and the device is quarantined from the network. However, if $L$ is connected through $R3$, then the access is not affected.
  

\begin{figure*}[!htb]
  \centering	
  \includegraphics[width=6.5in, scale=0.50]{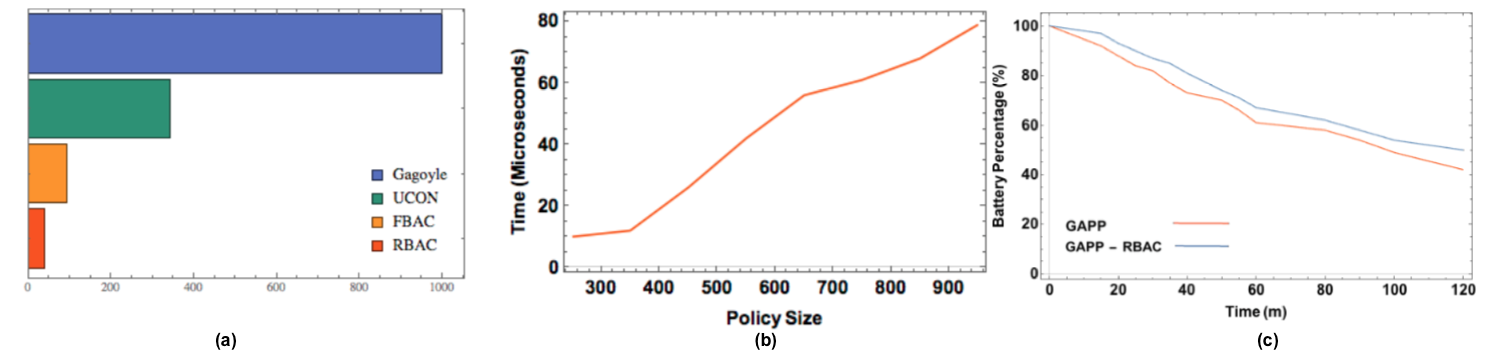}
  \caption{a) Comparison of Gargoyle with related work in in successfully protecting resources against insider scenarios implemented, b) Gargoyle's risk management average processing time as the policy size increases, and c) Gargoyle's mobile application (GAPP) energy overhead.}
  \label{architecture}
\end{figure*}


\subsection{Results Analysis} Gargoyle avoids a binary approach when making authorizations and depending on the context, a subset of functions may be disallowed for the data requester. As shown in Table III, in our simulated scenarios, more than half of access requests were granted despite detecting a threat. This proves the applicability of Gargoyle in the real-world context by enabling organizations to function securely even in the presence of threats. \par
\textbf{Baseline Comparison} Figure 5.a shows a comparison between RBAC, FBAC, \cite{lazouski2017stateful} and Gargoyle in successfully protecting resources against insider scenarios implemented. UCON model used in \cite{lazouski2017stateful} is successful in approximately 300 insider scenarios given that it ensures certain conditions are met before and post authorizations. However, the attributes used in this model are mainly limited to role, location and time. Hence, it fails to protect resources in about 70\% of the insider scenarios simulated. Moreover, one should note that any context information retrieved for this model are device driven and not extracted from the network as in the case of Gargoyle. This is a foundational difference between Gargoyle and related work such as \cite{lazouski2017stateful}. In the case of RBAC and FBAC, they both grant access authorizations depending on the DR's role within the organization. The granularity of FBAC manages limits the number of functions available to the user, and so, it can protect against about two times more insider scenarios despite not modeling the context properly. Hence, as also discussed in related work such as \cite{baracaldo2017g, crampton2010towards}, to protect organizations against insider threats, access control solutions must integrate context information as part of their decision process. \par

\textbf{Performance Metrics} Gargoyle extracts context information by passively intercepting network traffic. Hence, it does not have any impact on the network performance. On the user-side, however, Gargoyle must update the device's policy repository before access can be granted. As shown in Figure 5.b, even with more than 900 active policies for up to 90 active users, the average processing time is reasonable. In fact, Gargoyle's performance is acceptable when compared to related work such as \cite{baracaldo2017g} that have similar scope and methodology. Moreover, given that these are processed outside the user's device, the policy processing time and the system performance can be adjusted for larger organizations using services such as cloud computing and NFV. \par
Figure 5.c illustrates Gargoyle's mobile application (GAPP) performance. For this, we first fully charged our Samsung Galaxy S3 to 100 percent. Then, via a monkeyrunner \cite{monkey} script we run four file access requests every 2 minutes. For GAPP -- RBAC this was just accessing files as any other text editor by checking the user's role within the organization. For GAPP, this required retrieving policies from Gargoyle and enforcing them when two different functions were called for the file at random. Our results indicate that GAPP with FBAC does not incur a noticeable energy overhead. \par

\section{Related Work } \label{relatedwork}
The increasing adoption of mobile devices into organizations has motivated relatively sizeable literature discussing the importance of incorporating contextual factors such as location and time in access control \cite{bertino2011location}. For instance, \cite{shebaro2015context} proposed a modified version of the Android OS supporting context-based access control policies, which restrict applications from accessing specific data and resources based on the user context. However, this work along with much similar work before it requires the users to configure their own set of policies, it is essentially an extension of RBAC, and more importantly, context attributes are limited to location (absolute or relative) and time. Recently, \cite{lazouski2017stateful} proposes a framework to regulate the usage of data shared on mobile devices and regulates the right of using data continuously while access is in-progress. Specifically, Lazouski et al. build their framework on UCON model \cite{park2004ucon} rather than RBAC and retrieve policies from a remote server instead of relying on users for setup. Nevertheless, this solution also limits context to location and time while also relying on the user's device integrated sensors for retrieving them. The closest work to Gargoyle in terms of motivation and methodology is G-SIR \cite{baracaldo2017g}, which incorporates geo-social information as part of the AC system for insider threat mitigation. However, the attributes uses are completely different, and the framework assumes all information is already available. \par
Another category of related work is research aiming to detect insiders through the organization network traffic. Few proposals exist in the area and authors in \cite{liu2018detecting} and \cite{salem2008survey} have surveyed these solutions. For instance, ELICIT \cite{maloof2007elicit} detecting activities, such as searching, browsing, downloading, and printing, by monitoring the use of sensitive search terms, printing to a non-local printer and anomalous browsing activity. We regard these solutions as complimentary to Gargoyle as they are not designed as part of an access control solution. Furthermore, none of these solutions are built atop of SDN.

\section{Conclusion \& Future Work} \label{conclusion}
Gargoyle addresses the gap raised by \cite{crampton2010towards}, which specifically argues the requirement to integrate anomaly detection systems into organization's access control systems. In fact, to the best of our knowledge, Gargoyle is the first solution to evaluate the context of an access request using network-traffic extracted information such as the user's device capabilities, security-level, current and prior interactions with other devices, network connection status, and suspicious online activities. Furthermore, Gargoyle avoids a binary approach in access authorizations and by incorporating Function-based Access Control (FBAC), allows the customization of access policies into a set of functions depending on the trustworthiness of the context. Moreover, as a solution designed for organizations adopting SDN, it can also apply certain access restrictions at network-level and create a layered protection model. \par
Currently, our focus is to investigate Gargoyle's performance for larger networks when the context evaluation may lead to false positives. One possible solution to this would be to incorporate recent Machine Learning techniques for traffic analysis and adjusting Gargoyle's sensitivity when assessing context risk. Alternatively, we could also include host-level information and feedback into Gargoyle's PDP and validate the network-extracted information. In fact, integrating Gargoyle as complimentary to existing solutions may be the most practical approach for real-world deployment.

\section*{Acknowledgment}
We acknowledge the useful comments and insights provided by Prof. Yvo Desmedt and Dr. Sandra Scott-Hayward. All data collection and storage for this project was approved under HC15778 by UNSW Human Ethics Office.



\bibliographystyle{IEEEtran}
\bibliography{ref.bib}
%

\end{document}